\documentclass[twocolumn,superscriptaddress,nofootinbib,amsmath,amssymb,aps,floatfix]{revtex4-2}
\usepackage{graphicx,tikz} 
\usepackage[mode=buildmissing]{standalone} 
\usepackage{dcolumn}
\usepackage{bm}
\usepackage{physics}
\usepackage{xcolor}
\usepackage{siunitx}
\usepackage{soul}
\usepackage{comment}
\usepackage[bb=boondox]{mathalfa}
\usepackage{lipsum}
\usepackage{hyperref}

\newcommand{\perm}{\mathrm{perm}}

\begin{document}

\title{Quantum Tautochrone for photons and atoms}
\author{Georgios K. Simopoulos}
\email{gsimopoulos@physics.uoc.gr}
\affiliation{ITCP, Department of Physics, University of Crete, 70013, Heraklion, Greece}
\affiliation{Institute of Electronic Structure and Laser, Foundation for Research and Technology-Hellas (FORTH), P.O. Box 1527, 71110, Heraklion, Greece}
\author{Georgios E. Pavlou}
\affiliation{ITCP, Department of Physics, University of Crete, 70013, Heraklion, Greece}
\affiliation{Institute of Electronic Structure and Laser, Foundation for Research and Technology-Hellas (FORTH), P.O. Box 1527, 71110, Heraklion, Greece}
\author{Ioannis Kiorpelidis}
\affiliation{ITCP, Department of Physics, University of Crete, 70013, Heraklion, Greece}
\affiliation{Institute of Electronic Structure and Laser, Foundation for Research and Technology-Hellas (FORTH), P.O. Box 1527, 71110, Heraklion, Greece}
\author{Georgios A. Siviloglou}
\email{g.siviloglou@physics.uoc.gr}
\affiliation{ITCP, Department of Physics, University of Crete, 70013, Heraklion, Greece}
\affiliation{Institute of Electronic Structure and Laser, Foundation for Research and Technology-Hellas (FORTH), P.O. Box 1527, 71110, Heraklion, Greece}
\affiliation{Department of Physics, Southern University of Science and Technology, Shenzhen 518055, China}
\affiliation{International Shenzhen Quantum Academy, Shenzhen 518055, China}
\author{Konstantinos G. Makris}
\email{makris@physics.uoc.gr}
\affiliation{ITCP, Department of Physics, University of Crete, 70013, Heraklion, Greece}
\affiliation{Institute of Electronic Structure and Laser, Foundation for Research and Technology-Hellas (FORTH), P.O. Box 1527, 71110, Heraklion, Greece}

\begin{abstract}
The tautochrone problem, in which the travel time to a common point is independent of the starting position, has served as a foundational model in classical mechanics. Here we establish its quantum counterpart in engineered photonic waveguide arrays and atomic optical lattices, where nonuniform couplings drive photons and particles launched from different positions to converge simultaneously at a common site. Using higher-order correlation functions, we show that this quantum tautochrone effect holds for both bosons and fermions, with quantum statistics governing the character of the convergence: bosons converge onto the same site, while fermions, constrained by Pauli exclusion, gather at neighboring sites. For bosonic atoms, we further show how the tautochrone dynamics is shaped by lattice inhomogeneity, phase-space structure, and on-site interactions.
\end{abstract}

\date{\today}
\maketitle

\section{Introduction} 

The tautochrone phenomenon refers to the motion of a particle sliding under gravity along a curve such that its travel time to a target point is independent of its initial position on the curve. Huygens first demonstrated this property for frictionless motion along a cycloid, establishing one of the earliest and most celebrated results in classical mechanics \cite{Huygens1673}. Euler later expanded this framework to account for the friction encountered when moving through a resisting medium \cite{Euler1736}. In the context of classical mechanics, the equalization of travel times relies on shaping a potential. Since the tautochrone is a wave motion phenomenon, the question is if such a notion can be realized in other areas of classical and/or quantum physics. 

On the classical level, one relevant platform is that of integrated photonic waveguide arrays, since the coupling between neighboring channels can be designed at will in order to dictate how a wavepacket propagates. This precise control enables perfect state transfer, a protocol introduced for spin chains \cite{Bose2003,Christandl2004,Nikolopoulos2004,Vinet2012,Kay2010} and subsequently mapped onto optical waveguides
to steer classical wavepacket dynamics \cite{Longhi2010,Longhi2010_PRB,Bellec2012,
Chremmos2012}. Along these lines, photonic lattices with engineered coupling coefficients has been recently theoretically demonstrated
\cite{Kiorpelidis2026}. In particular, engineered inhomogeneity in the nearest-neighbor couplings, instead of a shaped potential, drives inputs launched at different positions to arrive simultaneously at a common site after a fixed propagation length, the tautochrone distance. 

A natural question, which is central in our work, is whether quantum states of light in such artificial systems can, in principle, support tautochrone dynamics, and how such an effect is imprinted on higher-order correlation functions, where quantum statistics produce features absent from classical intensity correlations~\cite{Bromberg2009,Lahini2010,Peruzzo2010,PerezLeija2013transport,Matthews2013,Swain2021}.

The platform of waveguide arrays, also extends to non-classical light, enabling the transport of path-entangled photons and photonic qubits \cite{PerezLeija2013,rechtsman2016topological,Chapman2016,Weimann2016,Preiss2015}. The same design freedom yields effective Hamiltonians with prescribed dynamics, from the Fock-space evolution of Glauber-Fock lattices \cite{PerezLeija2010,RodriguezLara2011,Keil2011,Longhi2009} to the disorder-induced localization of Anderson arrays \cite{Lahini2008}.

A second platform, is that of ultracold atoms in optical lattices, which offers the same capability of controlling the couplings, along with
tunable interactions and single-atom, statistics-resolved detection~\cite{GrossBloch2017}. These lattices have enabled the direct observation of coherent transport phenomena such as Bloch oscillations and Anderson localization \cite{BenDahan1996,Peik1997,Bloch2005, Roati2008}, Furthermore, through tunable on-site
interactions, these lattices enable access to many-body physics \cite{Bloch2008ManyBody,Bloch2008} including the superfluid--Mott insulator transition, repulsively bound pairs, and Hubbard-model dynamics \cite{Fisher1989,Greiner2002,Stoferle2004,Winkler2006,Kohl2005}, as well as engineered
effective spin Hamiltonians \cite{Duan2003}. This combination of coupling control,
interactions, and particle-resolved readout makes optical lattices a platform on which the quantum tautochrone can be probed for both bosons and fermions particles.

\begin{figure}
\centering
\includegraphics[width=1\linewidth]{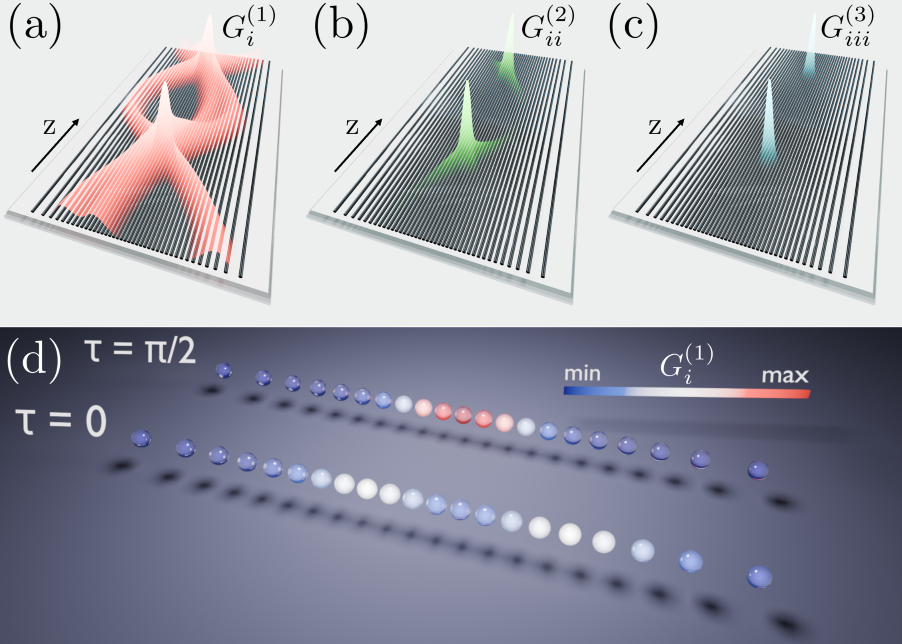}	
\caption{Quantum tautochrone in a nonuniform waveguide array for (a--c) three photons and (d) in optical lattices for two bosons. (a) At the tautochrone distance the average particle number $G_i^{(1)}$ reveals simultaneous convergence. (b,c) Second (third)-order autocorrelations $G_{ii}^{(2)}$ ($G_{iii}^{(3)}$) identify multi-particle coincidences. (d) Mean occupation number $G_i^{(1)}$ at $\tau=0$ and at the tautochrone time $\tau=\pi/2$.}
\label{fig:Tautochrone_lattice}
\end{figure}

In this work, we introduce the quantum tautochrone for photons and atoms in photonic waveguide arrays and atomic optical lattices, respectively. In these systems, quantum particles are described as spatially extended wavepackets. By engineering the lattice couplings, wavepackets launched from distinct positions can follow different trajectories yet still arrive at the same spatial region simultaneously. This is the definition of the tautochrone, which is evident in the engineered hopping dynamics, as well as in the presence of on-site interactions. We characterize this effect using higher-order correlation functions and Wigner functions. As shown schematically in Fig.~\ref{fig:Tautochrone_lattice}(a)--(c) for a waveguide array, the average photon number as well as the second- and third-order autocorrelations all reveal the tautochrone effect for three-photon excitations. In Fig.~\ref{fig:Tautochrone_lattice}(d), the mean occupation number for two atoms is shown at the initial and tautochrone times for an optical lattice. 

The paper is organized as follows. In Section~\ref{theoretical}, we establish the theoretical framework by defining the Hamiltonian of the problem and the relevant engineered coupling parameters that lead to the tautochrone. In Section~\ref{sec:Bosons_fermions} we demonstrate the tautochrone effect for two-particle bosonic and fermionic systems using correlation functions. In Section~\ref{sec:beyond_pst}, we establish the robustness of the tautochrone effect for a wide range of initial spatial widths and lattice inhomogeneities. In Section~\ref{sec:higher_correlations}, we explore the tautochrone using higher-order correlation functions as complementary observables. In Section~\ref{sec:wigner}, we employ Wigner functions to characterize the quantum state at the convergence region. In Section~\ref{sec:interactions}, we reveal how the tautochrone depends on the strength of on-site interactions. In Section~\ref{sec:conclusions} we summarize the main findings. In the Supplementary Material we derive the coupling profile that lead to tautochrone dynamics in the quantum realm and we present the calculation of the Wigner quasiprobability distribution as it appears in different locations in a one-dimensional lattice.

\section{Theoretical background}
\label{theoretical}

\begin{figure*}[t]
\centering
\includegraphics[width=1\linewidth]{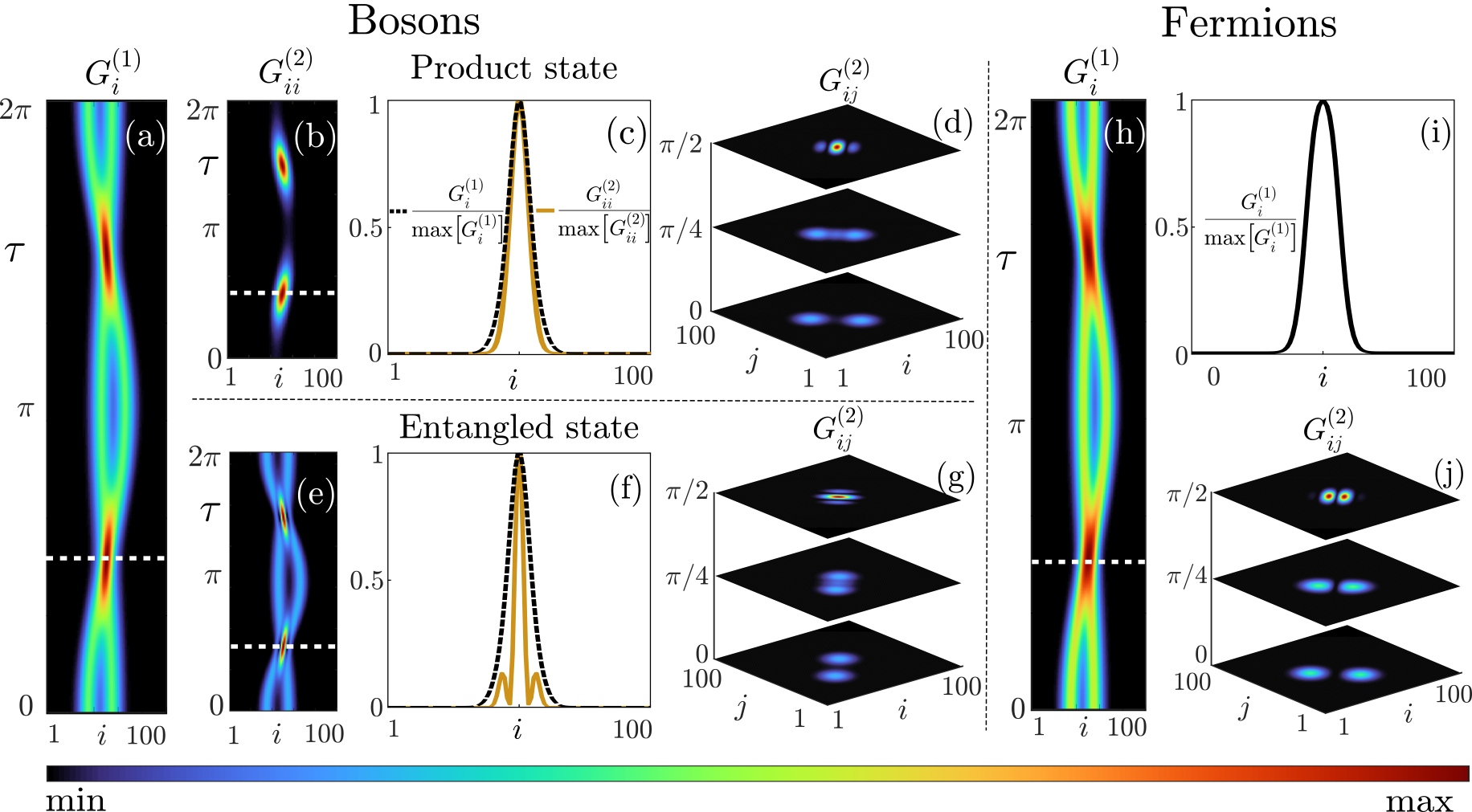}
\caption{Two-particle tautochrone dynamics in nonuniform lattices.
(a) $G_{i}^{(1)}$ for the product and superposition states with initial Gaussian wavepackets centered at sites $35$ and $55$ (Gaussian width $\sigma_m = 5$).
Panels (b)--(d) refer to the product state case, while panels (e)--(g) refer to the superposition case.
(b,e) $G_{ii}^{(2)}$. (c,f) Normalized density (dashed black) and second-order correlation profiles (solid golden yellow) at the tautochrone time $\tau_f=\pi/2$.
(d,g) $G_{ij}^{(2)}$ at $\tau=0$, $\pi/4$, and $\pi/2$. (h--j) Similar to (a,c,d), respectively, for the fermionic product state. Minimum values: all zero. Maximum values: (a) $0.1843$, (b,d,e,g) $0.0268$, (h) $0.1461$, (j) $0.0169$.
}
\label{fig:2particles}
\end{figure*}

In this Section we introduce a common theoretical framework to explore quantum tautochrone dynamics in nonuniform lattices for photons or atoms. The system consists of $N$ non-interacting spinless quantum particles on a one-dimensional lattice of $L$ sites. The corresponding state evolves according to
\begin{equation}\label{eq:Schrodinger}
    i \frac{d}{dt} \lvert \psi \rangle = \hat{H} \lvert \psi \rangle ,
\end{equation}
where $\hbar = 1$ and $t$ denotes the evolution time. In photonic waveguide arrays, the same evolution equation applies with the replacement $t\rightarrow z$, where $z$ is the propagation distance along the array. For simplicity, in what follows we refer to the tautochrone time throughout, with the understanding that this corresponds to the tautochrone distance in the case of waveguide arrays.

The Hamiltonian that governs the evolution of the system is given by
\begin{equation}
    \hat{H} = -\sum_{i=1}^{L-1} J_i \left( \hat{a}_{i+1}^\dagger \hat{a}_i +  \hat{a}_{i}^\dagger \hat{a}_{i+1} \right), \label{eq:Hamiltonian}
\end{equation}
where $\hat{a}_i^\dagger$ ($\hat{a}_i$) creates (annihilates) a particle at site $i$, $J_i$ is the nearest-neighbor tunneling amplitude, and $\hat{n}_i=\hat{a}_i^\dagger\hat{a}_i$ is the local number operator. For bosons, $[\hat a_i,\hat a_j^\dagger]=\delta_{ij}$, whereas for fermions, $\{\hat a_i,\hat a_j^\dagger\}=\delta_{ij}$.

The tunneling amplitude is engineered to produce the desired tautochrone behavior: wavepackets launched from different positions reach the same spatial region simultaneously. The coupling profile can be constructed by requiring the center of the wavepacket to follow harmonic oscillations along the lattice. The position of the wavepacket in the lattice is tracked by the expectation value $\langle x\rangle$, which was shown to oscillate with frequency $2\omega$ in classical waveguide arrays \cite{Kiorpelidis2026}. In the quantum case, the mean particle number is expected to exhibit similar oscillatory dynamics (Supplementary Note 1). Thus, we obtain the same nonuniform coupling profile
\begin{equation}
\label{eq:J_profile}
J_i = \omega \sqrt{ C^2 - \left(i-\frac{L}{2}\right)^2 },
\end{equation}
where $C\geq L/2$ controls the strength of the coupling inhomogeneity across the lattice and sets the amplitude of the oscillations around the lattice center ($L/2$). Choosing $C=L/2$, the profile becomes the $J_x$ lattice used for perfect state transfer \cite{Christandl2004,Longhi2010,PerezLeija2013}, for which Eq.~\eqref{eq:J_profile} reduces to

\begin{equation}
\label{eq:J_profile_perfect}
J_i = \omega \sqrt{ i(L-i) }.
\end{equation}

Since the wavepackets used for the tautochrone dynamics are spatially extended, we describe them using a single-particle creation operator 
\begin{equation}\label{eq:Operators}
\hat{A}_m^\dagger=\sum_{i=1}^L\phi_m(i)\hat{a}_i^\dagger,
\end{equation}
where $\phi_m(i)$ is the spatial profile of the normalized single-particle on site $i$ for the $m$-th input mode. Without loss of generality, throughout this work for the $m$-th mode we use a Gaussian wavepacket
\begin{equation}
\phi_m(i) = \mathcal{N}\sum_{l=1}^{N_{\text{lobes}}(m)}K_m^{(l)}\exp\left[-\frac{(i-x_m^{(l)})^2}{4\sigma_m^2}\right], \label{eq:lobes}
\end{equation}
where $l$ is the Gaussian lobe out of the total $N_{\text{lobes}}(m)$, $x_m^{(l)}$ is the initial center, $\sigma_m$ is the spatial width, $K_m^{(l)}$ is the amplitude of each lobe and $\mathcal{N}$ is the normalization constant. In all sections we use $K_{m}^{(l)}=1$ and $N_{\text{lobes}}(m)=1$, dropping the superscript in this case, except in one case discussed in Section~\ref{sec:wigner}, where $N_{\text{lobes}}(m)=2$.

Using the wavepackets defined in Eq.~\eqref{eq:lobes}, we construct the initial state through Eq.~(\ref{eq:Operators})  acting on the vacuum state, evolve it according to Eq.~\eqref{eq:Schrodinger}, and characterize its dynamics using normally ordered correlation functions \cite{Glauber1963},
\begin{equation}
\begin{aligned}
G_i^{(1)}(\tau)&=\langle \hat a_i^\dagger \hat a_i\rangle=\langle \hat n_i\rangle,\\
G_{ij}^{(2)}(\tau)&=\langle \hat a_i^\dagger \hat a_j^\dagger \hat a_j \hat a_i\rangle,\\
G_{ijk}^{(3)}(\tau)&=\langle \hat a_i^\dagger \hat a_j^\dagger \hat a_k^\dagger \hat a_k \hat a_j \hat a_i\rangle,
\end{aligned}
\end{equation}
with the expectation value taken in the time-evolved state. We use the dimensionless time $\tau=2\omega t$. Here, $G_i^{(1)}$ is the mean occupation density, $G_{ij}^{(2)}$ is the joint detection weight at sites $i$ and $j$, and $G_{ijk}^{(3)}$ captures simultaneous three-particle coincidences. These multi-particle coincidences provide direct evidence for the quantum tautochrone and have no counterpart in classical laser beam dynamics~\cite{Kiorpelidis2026}.

\section{The quantum tautochrone for bosons and fermions}
\label{sec:Bosons_fermions}

Using the coupling profile $J_i$ introduced in the previous section, we demonstrate that the quantum tautochrone effect emerges broadly across bosonic and fermionic wavepacket states, leaving its signature in both first- and second-order correlations. We begin with the simplest case of two non-interacting particles, deferring the role of interactions to a later section.

We consider a system in which the two particles are initially delocalized, as described by the Gaussians in Eq.~\eqref{eq:lobes}. Figure~\ref{fig:2particles} illustrates their evolution. Panels (a)--(g) correspond to bosons (which can be photons or bosonic atoms), whereas panels (h)--(j) correspond to fermionic atoms. For the two-particle product state, we use
\begin{equation}\label{eq:product}
    \lvert \psi_{\mathrm{prod}}^{(2)}(0) \rangle = \frac{1}{\sqrt{1 \pm \lvert S_{12} \rvert^2}} \hat{A}_1^\dagger \hat{A}_2^\dagger \lvert 0 \rangle,
\end{equation}
where the upper (lower) sign applies to bosons (fermions) and $S_{nm}=\sum_{i=1}^{L}\phi_{n}^{*}(i)\phi_{m}(i)$ is the wavepacket overlap. This state consists of one particle for each of the two Gaussian wavepackets. 

By contrast, in the following state, both particles occupy either the first or the second Gaussian wavepacket:
\begin{equation}\label{eq:entangled}
    \lvert \psi_{\mathrm{sup}}^{(2)}(0) \rangle = \frac{1}{2\sqrt{\left(1 + \text{Re}(S_{12}^2)\right)}} \left( \hat{A}_1^{\dagger 2} + \hat{A}_2^{\dagger 2} \right) \lvert 0 \rangle.
\end{equation}
This state is a normalized superposition of two-boson occupations in two partially overlapping Gaussian wavepackets. In the limit of zero overlap, $S_{12}=0$, the state becomes maximally entangled. Such a state can be readily prepared for photons using a Hong–Ou–Mandel setup, with the two output ports shaped and guided, with minimal overlap, into the tautochrone lattice \cite{Hong1987}.

In the system, we consider $L=100$ lattice sites, $C=L/2$, and the two particles have $\sigma_m = 5$ each. The first-order correlator $G_i^{(1)}$ measures the mean occupation number at site $i$. Therefore, it captures the simultaneous convergence of the two wavepackets at the central region at the tautochrone time $\tau_f=\pi/2$ as seen in Fig.~\ref{fig:2particles}(a).

However, to demonstrate the quantum tautochrone effect, it is essential to consider the second-order correlators. In Fig.~\ref{fig:2particles}(b) we present the evolution of the diagonal $G^{(2)}_{ii}$, which quantifies the coincidence of two particles on site $i$, for the product state $\lvert \psi_{\mathrm{prod}}^{(2)}(\tau)\rangle$. Figure~\ref{fig:2particles}(c) displays the normalized first- and second-order profiles at the tautochrone time. These features originate from interference between the two independently-evolving single-particle wavepackets. To make this explicit and complement our numerical results, we derive semi-analytical expressions for the correlators. Let $\phi_m(i,\tau)$ denote the spatial profile on site $i$ at time  $\tau$ of mode $m$ evolved in isolation, with $\phi_m(i,0)$ the initial profile; we stress that $\phi_m(i,\tau)$ is merely a label for the propagated wavepacket and does not imply that its functional form is preserved under evolution.  The quantum correlators then follow directly from these single-particle profiles. For the product state, the first-order correlator is \cite{Mandel_Wolf_1995}
\begin{equation}
\begin{aligned}
G_i^{(1)}(\tau)
&=\frac{1}{1\pm |S_{12}|^2}\Bigl[
|\phi_1(i,\tau)|^2+|\phi_2(i,\tau)|^2 \\
&\qquad\quad{}
\pm 2\,\mathrm{Re}\bigl[S_{21}\phi_1^*(i,\tau)\phi_2(i,\tau)\bigr]
\Bigr],
\label{eq:analyticalG1}
\end{aligned}
\end{equation}
whereas the second-order correlator between sites $i$ and $j$ is \cite{Kaufman2018}
\begin{equation}
\begin{aligned}
    G^{(2)}_{ij}(\tau) &= \frac{\big|\phi_1(i,\tau)\phi_2(j,\tau)\pm\phi_2(i,\tau)\phi_1(j,\tau)\big|^2}{1\pm |S_{12}|^2}.
    \label{eq:analyticalG2}
\end{aligned}
\end{equation}
Here and in Eq.~\eqref{eq:analyticalG1}, the upper (lower) sign corresponds to bosons (fermions). Equation~\eqref{eq:analyticalG2} describes both diagonal and off-diagonal correlations. On the diagonal, it reduces to
\begin{equation}
G^{(2)}_{ii}(\tau)=
\begin{cases}
\displaystyle \frac{4|\phi_1(i,\tau)|^2|\phi_2(i,\tau)|^2}{1+|S_{12}|^2}, & \text{bosons},\\[6pt]
0, & \text{fermions}.
\end{cases}
\label{eq:analyticalG2diagonal}
\end{equation}

Thus, for the bosonic product state shown in Fig.~\ref{fig:2particles}(b,c), $G_i^{(1)}$ is linear in the single-particle density contributions, while $G^{(2)}_{ii}$ is proportional to their product. When the two wavepackets have comparable Gaussian profiles near the convergence region, this product has a narrower envelope than the first-order density. For the fermionic case, $G^{(2)}_{ii}$ vanishes identically, as expected from the Pauli exclusion principle.

The product state retains pronounced off-diagonal weight as shown in Fig.~\ref{fig:2particles}(d), corresponding to cases in which the particles occupy different sites. At the tautochrone time, this off-diagonal weight is strongly reduced, and a single Gaussian lobe at the convergence site becomes the dominant feature.

The first-order density of the two-mode bosonic superposition of Eq.~\eqref{eq:entangled} is
\begin{equation}
\begin{aligned}
G_i^{(1)}(\tau)
&=\frac{1}{1+\mathrm{Re}[S_{12}^2]}\Bigl[
|\phi_1(i,\tau)|^2+|\phi_2(i,\tau)|^2 \\
&\qquad\quad{}
+ 2\,\mathrm{Re}\bigl[S_{12}\phi_1^*(i,\tau)\phi_2(i,\tau)\bigr]
\Bigr].
\label{eq:analyticalG1_NooN}
\end{aligned}
\end{equation}
For the real Gaussian input profiles considered here, the overlap $S_{12}$ is real, and therefore $S_{21}=S_{12}$ and $\mathrm{Re}[S_{12}^{2}]=|S_{12}|^2$. Equation~\eqref{eq:analyticalG1_NooN} then coincides with the bosonic branch of Eq.~\eqref{eq:analyticalG1}, explaining why the product and superposition states have the same site-resolved density in Fig.~\ref{fig:2particles}(a). However, this apparent equivalence is only first-order: the second-order correlation $G_{ii}^{(2)}$ clearly separates the two states, as shown in Fig.~\ref{fig:2particles}(e). At the tautochrone time, as shown in Fig.~\ref{fig:2particles}(f), the second-order autocorrelator $G^{(2)}_{ii}$ reaches its maximum value at the middle site, illustrating the self-interference of the
two particles. For the entangled (bosonic) state, this correlator is given by
\begin{equation}
    G_{ij}^{(2)}(\tau)
    =
    \frac{
        \left|
        \phi_1(i,\tau)\phi_1(j,\tau)
        +
        \phi_2(i,\tau)\phi_2(j,\tau)
        \right|^2
    }
    {
        1+\mathrm{Re}[S_{12}^2]
    }.
	\label{eq:N2G2_NOON}
\end{equation}
The superposition state concentrates its correlation weight closer to the diagonal in the second-order cross-correlation shown in Fig.~\ref{fig:2particles}(g), indicating an enhanced probability of joint occupation of the same spatial region.

For fermions, the tautochrone lattice still drives the two wavepackets toward convergence at the same spatial region, yet the Pauli exclusion principle forbids two identical fermions from occupying the same lattice site, precluding the concentration of two-particle correlation weight on the diagonal elements $G_{ii}^{(2)}$. In what follows, we examine how the tautochrone effect manifests for fermions under this constraint. The two fermions are described by the product state in Eq.~\eqref{eq:product}, as shown in Figs.~\ref{fig:2particles}(h)--\ref{fig:2particles}(j). Since the initial wavepackets have a finite overlap, $S_{12}=\sum_{i=1}^{L}\phi_{1}^{*}(i)\phi_{2}(i)\approx 0.14$, the fermionic state also produces a first-order density that differs from the corresponding bosonic case, as seen by comparing Figs.~\ref{fig:2particles}(a) and \ref{fig:2particles}(h). Unlike $G_{ii}^{(2)}$, which vanishes identically for fermions as a consequence of $\hat{a}_i^{\dagger 2}=0$, the first-order density $G_i^{(1)}$ is nonzero wherever either single-particle wavefunction is present. This mean occupation number shows that the wavepackets converge at $\tau_f=\pi/2$, with the Gaussian lobe in the cross section of Fig.~\ref{fig:2particles}(i) extending to more lattice sites than in the corresponding bosonic case, because the initial wavepackets are partially overlapping. This same exclusion is directly visible in Fig.~\ref{fig:2particles}(j), where $G_{ii}^{(2)}=0$ produces a zero along the diagonal at every site. As a result, only bosons exhibit true two-particle convergence at the same lattice site, while fermions instead gather at sites adjacent to those at the middle of the lattice. As we will see in Section~\ref{sec:interactions}, interactions can have a similar influence on the quantum tautochrone.

\begin{figure}
\centering
\includegraphics[width=1\linewidth]{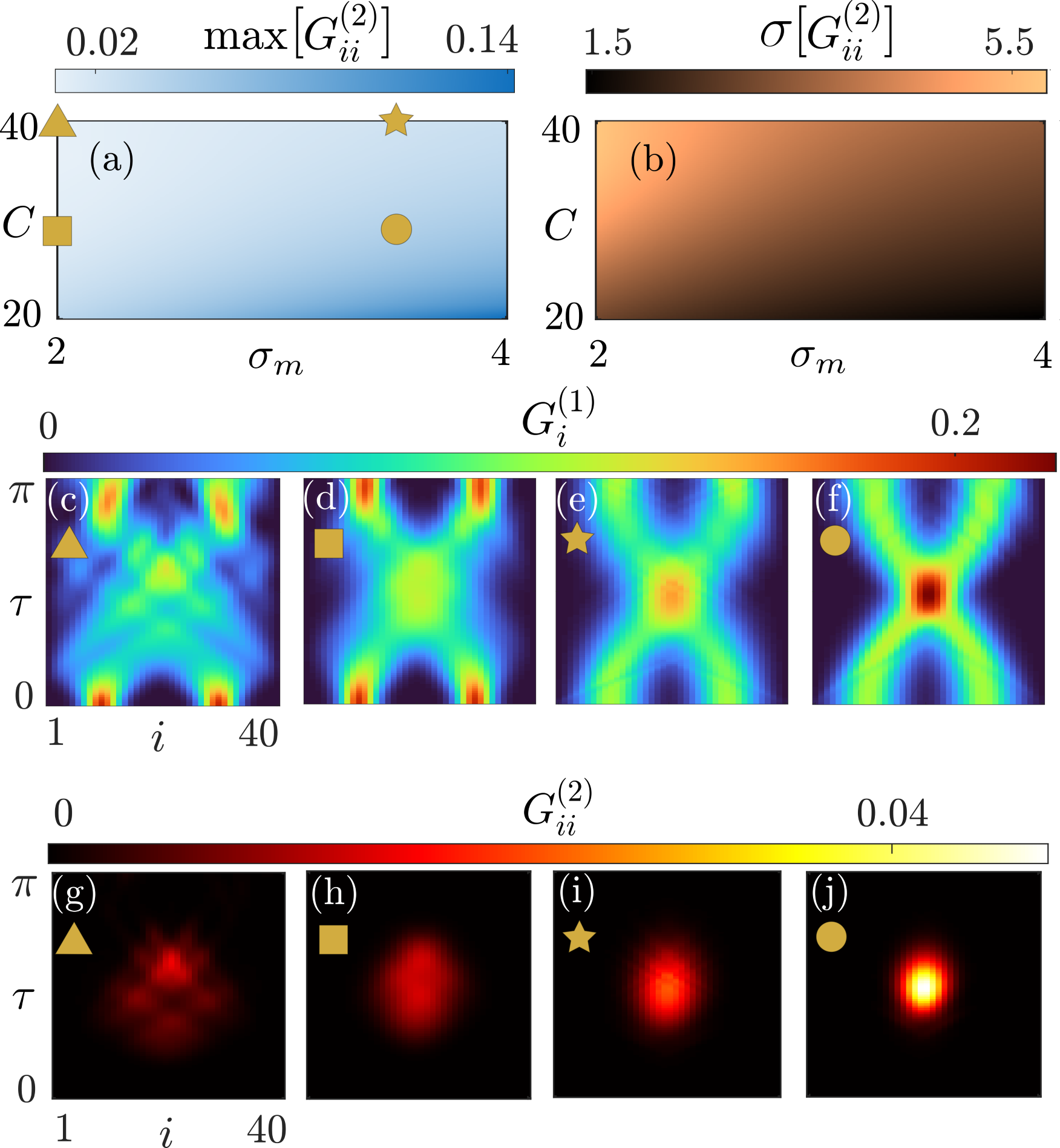}
\caption{Effect of coupling inhomogeneity and Gaussian width. All panels refer to the evolution of two bosons placed at lattice sites $10$ and $30$, respectively.
(a) Maximum value of the second-order correlation function, $\max[G_{ii}^{(2)}]$, evaluated at the tautochrone time $\tau_f=\pi/2$, as a function of the inhomogeneity parameter $C$ and the Gaussian width $\sigma_m$.
(b) Standard deviation $\sigma[G^{(2)}_{ii}]$ of the second-order correlation function at the tautochrone time.
The markers indicate the parameter combinations used in panels (c)--(j).
(c)--(f) Evolution of the site-resolved density $G_{i}^{(1)}$ for four choices of $C$ and $\sigma_m$: (c) $C=40, \sigma_m=2$; (d) $C=28, \sigma_m=2$; (e) $C=40, \sigma_m=3.6$; and (f) $C=28, \sigma_m=3.6$.
(g)--(j) Corresponding evolution of the local second-order correlation function $G_{ii}^{(2)}$.
}
\label{fig:C_N_2}
\end{figure}

\section{Robustness of the quantum tautochrone}
\label{sec:beyond_pst}
The sharpness with which the quantum tautochrone emerges in the correlation functions depends on two competing parameters: the spatial variation of the couplings and the width of the initial wavepackets. In Fig.~\ref{fig:C_N_2}, we illustrate the evolution of two bosons, initially placed symmetrically around the middle of a lattice of size $L=40$, with each particle described by a Gaussian wavepacket of width $\sigma_m$. This configuration allows us to vary both the coupling inhomogeneity parameter $C$ and the Gaussian width $\sigma_m$, in order to determine how the convergence at the location of the tautochrone is affected.

\begin{figure*}
\centering
\includegraphics[width=1\linewidth]{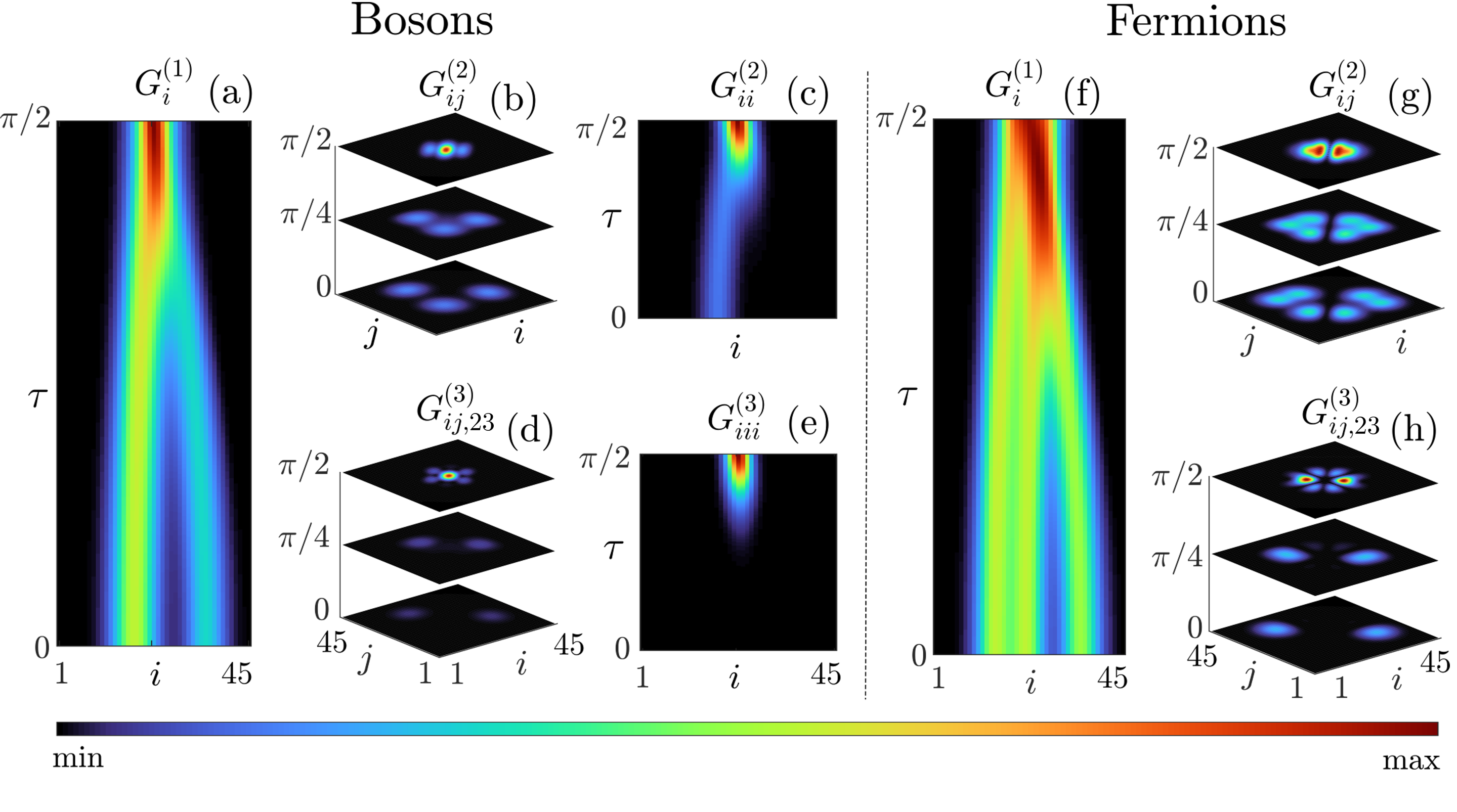}
\caption{Three-particle tautochrone dynamics.
Initial states consist of three particles centered at sites $17$, $20$, and $35$ with the appropriate $\sigma_m$ for a lattice with $L=45$. The initial state is $|\psi_{\mathrm{prod}}^{(3)}\rangle$ for bosons and fermions. 
(a) Mean occupation $G^{(1)}_{i}(\tau)$ for bosons. 
(b) Second-order correlators $G_{ij}^{(2)}$ for bosons. 
(c,e) Bosonic autocorrelators: (c) $G_{ii}^{(2)}$ and (e) $G_{iii}^{(3)}$. 
(d) Third-order correlators $G_{ijk}^{(3)}$ ($k=23$) for bosons. (f--h) Similar to (a,b,d), but for fermions. Minimum values: all zero. Maximum values: (a) $0.419$, (b,c) $0.182$, (d,e) $0.042$, (f) $0.267$, (g) $0.059$, (h) $0.011$. Correlations presented in the panels evolve up to the tautochrone time $\tau_f=\pi/2$.} 
\label{fig:three_particle_tautochrone}
\end{figure*}

To show quantitatively the two-particle convergence at the tautochrone region, we present in Fig.~\ref{fig:C_N_2}(a) the maximum value of the second-order correlations, $\max[G_{ii}^{(2)}]$, evaluated at the tautochrone time $\tau_f = \pi/2$, as a function of $C$ and $\sigma_m$. Increasing $C$ leads to a reduction of $\max[G_{ii}^{(2)}]$. Increasing the oscillation amplitude $C$ spreads the same two-particle correlation weight over a larger region of the lattice. In contrast, a larger initial width $\sigma_m$ increases $\max[G_{ii}^{(2)}]$, confirming that broader wavepackets lead to better focusing. To quantify the spatial extent of the wavepackets at the tautochrone, we present in Fig.~\ref{fig:C_N_2}(b) the standard deviation $\sigma[G_{ii}^{(2)}]$ at the tautochrone time. Increasing $C$ results in a larger standard deviation, while increasing $\sigma_m$ reduces it. 

The decrease in $C$ and the increase in $\sigma_m$ lead to higher correlation values, which correspond to a stronger signature of the tautochrone. Figures~\ref{fig:C_N_2}(c)--\ref{fig:C_N_2}(f) illustrate this behavior through the time evolution of the site-resolved density $G_{i}^{(1)}$. A more direct confirmation is provided in Figs.~\ref{fig:C_N_2}(g)--\ref{fig:C_N_2}(j), where we show the corresponding second-order correlation function $G_{ii}^{(2)}$ for four representative choices of $C$ and $\sigma_m$, also indicated in Fig.~\ref{fig:C_N_2}(a): $C=40, \sigma_m=2$ (triangle marker), $C=28, \sigma_m=2$ (square marker), $C=40, \sigma_m=3.6$ (star marker), $C=28, \sigma_m=3.6$ (circle marker). For large $C$ combined with small $\sigma_m$ (triangle marker), the wavepackets spread significantly during the evolution, washing out any clear tautochrone signature. Reducing $C$ while keeping $\sigma_m$ small (square marker) is already sufficient to reverse this trend: the two particles approach the lattice center and the tautochrone becomes noticeable. A comparable improvement follows from increasing $\sigma_m$ instead, while keeping $C$ large (star marker), where the correlation dynamics show a clear convergence at the tautochrone. The two effects reinforce each other when $C$ is reduced and $\sigma_m$ increased simultaneously (circle marker), yielding a stronger and more localized correlation peak at the tautochrone time.

\section{Higher-order correlations of the quantum tautochrone}
\label{sec:higher_correlations}

In the current section we show the significance of higher-order correlations when identifying the simultaneous many-particle convergence at the tautochrone. In Fig.~\ref{fig:three_particle_tautochrone} we present the dynamics of a three-boson product state and a three-fermion product state. The structure of the initial condition is described by 

\begin{equation}\label{eq:product3}
    \lvert \psi_{\mathrm{prod}}^{(3)}(0) \rangle = \frac{1}{\mathcal{M}}\hat{A}_1^\dagger \hat{A}_2^\dagger \hat{A}_3^\dagger \lvert 0 \rangle,
\end{equation}
where the normalization constant is calculated as $\mathcal{M}= \sqrt{1 \pm \lvert S_{12} \rvert^2\pm \lvert S_{13} \rvert^2\pm \lvert S_{23} \rvert^2 + 2\text{Re}(S_{12}^{*}S_{13}S_{23}^{*})}$, where the upper (lower) signs apply to bosons (fermions).
The Gaussian width $\sigma_m=\frac{1}{\sqrt{2}}\left[C^2-\left(x_m-\frac{L+1}{2}\right)^2\right]^{1/4}$ ensures that the wavepackets maintain the same width throughout propagation \cite{Kiorpelidis2026}. We choose this value of $\sigma_m$ for simplicity, so that focusing effects do not play a significant role in the tautochrone dynamics.

\begin{figure*}[t]
	\centering
	\includegraphics[width=1\linewidth]{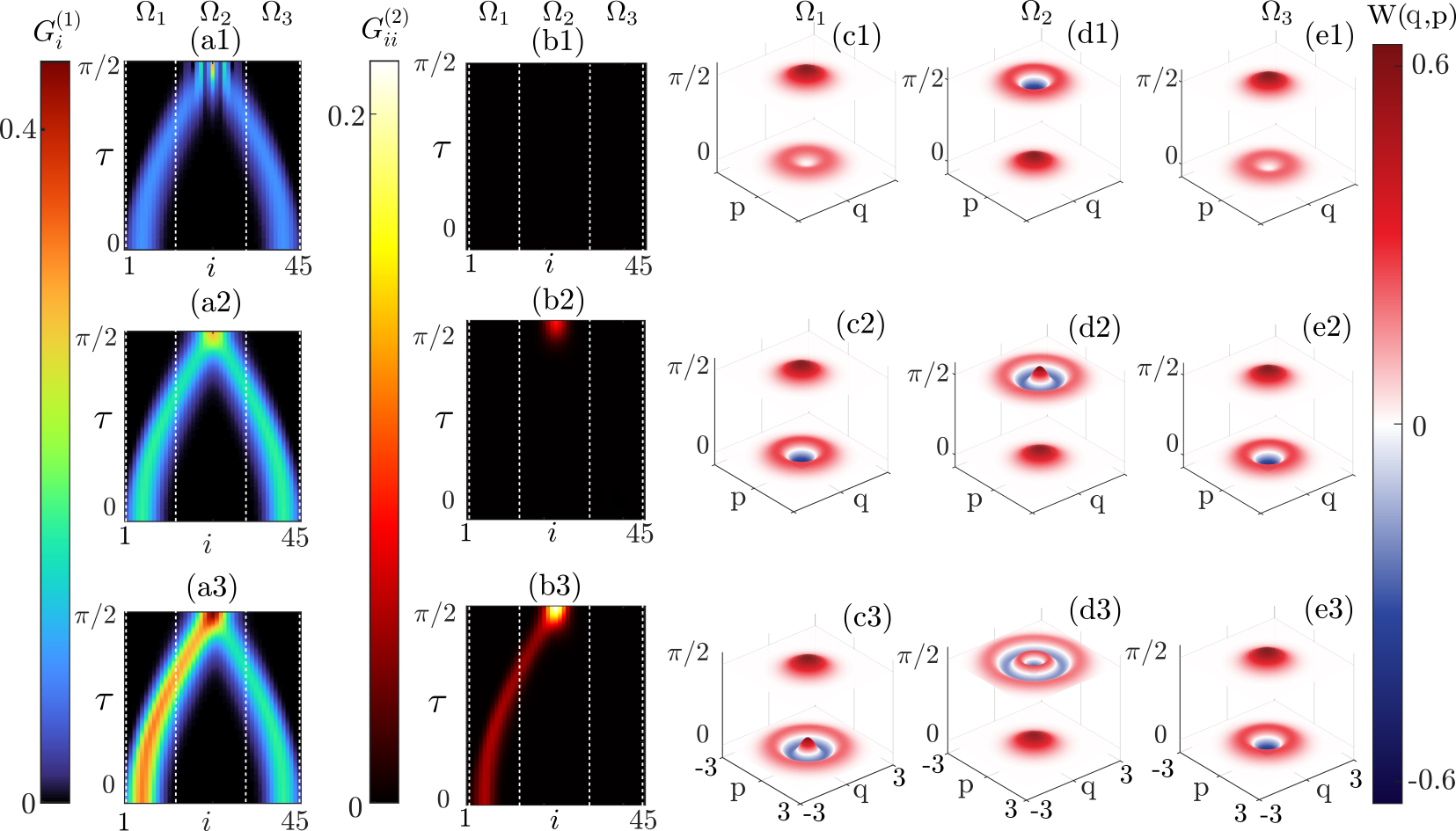}
	\caption{Correlations and Wigner functions for bosons.
		Dynamics for three Gaussian initial states ($\sigma_m=2.5$) centered at lattice sites $5$ and $41$: (1) a delocalized single-boson Fock state with superposition lobes at both sites, (2) a pure Fock-Fock state with one boson per site, and (3) a two-boson Fock state at $5$ and one-boson Fock state at $41$.
        (a1)--(a3) Mean occupation number $G^{(1)}_i(\tau)$.
        (b1)--(b3) Local second-order autocorrelators $G^{(2)}_{ii}(\tau)$, with dashed lines referring to the spatial measurement windows.
        (c1)--(e3) Wigner functions $W_{\Omega_i}(q,p)$ computed at $\tau=0$ and $\tau_f=\pi/2$ for the first (c), second (d), and third (e) measurement windows.}
	\label{fig:wigner_dynamics}
\end{figure*}

To illustrate the trajectories of the three bosons we show the first-order dynamics $G^{(1)}_i(\tau)$ in Fig.~\ref{fig:three_particle_tautochrone}(a). At $\tau=0$, all three particles are centered in different locations, $x_1 = 17$, $x_2 = 20$, and $x_3 = 35$, while at the tautochrone time they converge at the middle site. The same observation takes place in Fig.~\ref{fig:three_particle_tautochrone}(b) at the calculation of the second-order correlators $G^{(2)}_{ij}$. There is some spatial overlap between two of them, which explains the existence of lobes before the tautochrone time. The diagonal of the second-order correlators $G^{(2)}_{ij}$ in Fig.~\ref{fig:three_particle_tautochrone}(b), which is exactly the autocorrelator $G^{(2)}_{ii}(\tau)$ presented in Fig.~\ref{fig:three_particle_tautochrone}(c), stays nonzero long before the tautochrone. As we move towards the tautochrone time all three lobes in Fig.~\ref{fig:three_particle_tautochrone}(b) approach the middle site. 

While the second-order correlator captures pairwise overlap, the simultaneous convergence of all three particles is encoded in $G^{(3)}_{ijk}(\tau)$. For the product state in Eq.~\eqref{eq:product3}, the bosonic third-order correlator is \cite{Kaufman2018}
\begin{equation}
    G^{(3)}_{ijk}(\tau) = \frac{1}{\mathcal{M}^2}\left|\,\perm\!\begin{pmatrix}
        \phi_1(i,\tau) & \phi_2(i,\tau) & \phi_3(i,\tau)\\
        \phi_1(j,\tau) & \phi_2(j,\tau) & \phi_3(j,\tau)\\
        \phi_1(k,\tau) & \phi_2(k,\tau) & \phi_3(k,\tau)
    \end{pmatrix}\right|^2 ,
\label{eq:N3G3}
\end{equation}
where, the indices $i$, $j$, and $k$ correspond to the detection sites, while the subscripts $m=1,2,3$ to the three initial wavepackets. Bosonic symmetrization gives the matrix permanent, while for fermions, the permanent is replaced by the determinant. We present in Fig.~\ref{fig:three_particle_tautochrone}(d) the most physically relevant slice of this three-dimensional correlator: the slice for $k=23$, with one detection site fixed at the lattice center. Before the tautochrone time, this slice exhibits two lobes because the particles remain separated and the central site is unoccupied. The bosons approach the location of the tautochrone in a similar manner as seen in the two-particle case in Fig.~\ref{fig:2particles}. The expected number of simultaneous three-boson combinations is maximal at the tautochrone time $\tau_f = \pi/2$. A clearer picture of the tautochrone can be obtained when observing the Gaussian lobe in Fig.~\ref{fig:three_particle_tautochrone}(e), because it filters out unwanted combinations of particles focusing attention on the Gaussian lobe at the tautochrone time.

At the tautochrone time, the first-order correlations are nonzero for fermions, as shown in Fig.~\ref{fig:three_particle_tautochrone}(f). This is not the case, however, for the higher-order correlators. In the fermionic second-order correlator $G^{(2)}_{ij}$, shown in Fig.~\ref{fig:three_particle_tautochrone}(g), the Pauli exclusion principle gives rise to six distinct lobes prior to the tautochrone time, corresponding to the six possible pairwise particle configurations. The third-order correlator in Fig.~\ref{fig:three_particle_tautochrone}(h), however, exhibits only two lobes, similarly to the bosonic case. Not only do the autocorrelations $G^{(2)}_{ii}(\tau)$ and $G^{(3)}_{iii}(\tau)$ vanish, but so do all third-order correlations in which any two of the three indices $i,j,k$ coincide with the middle site. While the three fermions are, similarly to the bosons, driven toward the spatial center by the tautochrone dynamics, the Pauli exclusion principle strictly forbids their overlap. We note that the order of the correlation function required to capture the effect scales with the number of particles in the system.

\section{Wigner function characterization of the tautochrone dynamics for bosons}
\label{sec:wigner}

The correlation functions discussed so far reveal whether particles converge at the tautochrone time. An alternative way of describing such an effect is based on Wigner functions \cite{Banaszek1999,Royer1977,Cahill1969}. This phase-space representation is informationally equivalent to the density matrix, encoding both the populations and the coherences of the selected mode, and is therefore useful for revealing nonclassical features such as phase-space negativity where present.

More specifically, we evaluate the Wigner functions in selected spatial partitions $\Omega_i$, which act as detection regions. By removing the contribution of all sites outside each window, $W_{\Omega_i}(q,p)$ characterizes the quantum state formed near the convergence region. As a natural choice to demonstrate the spatial dependence of the quantum state, we take the first, second, and third spatial partitions depicted in Fig.~\ref{fig:wigner_dynamics} (here denoted as $\Omega_1 = \{1, \dots, 13\}$, $\Omega_2 = \{14, \dots, 31\}$, and $\Omega_3 = \{32, \dots, 45\}$), which together cover the entire lattice. We calculate the probability $P_{n}(\Omega_i)$ of finding exactly $n$ particles in window $\Omega_i$, by projecting the state onto the particle-number sectors of the chosen spatial window $\Omega_i$ (our general methodology for arbitrary states is presented in Supplementary Note 2). For compactness, we denote the particle-number distribution in a window by

\begin{equation}
	\mathbf{P}(\Omega_i)=(P_0(\Omega_i),P_1(\Omega_i),P_2(\Omega_i), P_3(\Omega_i))\label{eq: Probabilities_Wigner}.
\end{equation}

To this end, we construct a Wigner function $W_{\Omega_i}(q,p)$ from the occupation-number distribution $P_{n}(\Omega_i)$. Here, $q$ and $p$ are the dimensionless phase-space coordinates used to evaluate the Wigner function, related to the complex amplitude by $\alpha=(q+ip)/\sqrt{2}$. They are the standard quadratures of the single collective mode defined by window $\Omega_i$ (see Supplementary Note 2). We recall that a Fock state $\ket{n}$ has the Wigner function $W_n(q,p) \propto (-1)^n L_n(4|\alpha|^2)\,e^{-2|\alpha|^2}$, whose negativity and radial structure are set by the Laguerre polynomial $L_n$.

We consider three initial configurations, all formed by Gaussian lobes centered at sites 5 and 41, with $\sigma_m = 2.5$ for every particle index $m$. In the first configuration, shown in Fig.~\ref{fig:wigner_dynamics}(a1)--(e1), a single particle is prepared in a superposition of the two lobes, described by Eq.~\eqref{eq:lobes} with $N_{\text{lobes}}(1) = 2$, $x_1^{(1)}= 5$, and $x_1^{(2)}= 41$. In the second configuration, shown in Fig.~\ref{fig:wigner_dynamics}(a2)--(e2), two particles are initialized as one particle in each lobe, so $N_{\text{lobes}}(1) = 1$, $x_1= 5$, and $x_2= 41$. In the third configuration, shown in Fig.~\ref{fig:wigner_dynamics}(a3)--(e3), three particles are used: two occupy the lobe centered at site 5 and one occupies the lobe centered at site 41, thus $N_{\text{lobes}}(1) = 1$, $x_1=x_2= 5$, and $x_3= 41$. The columns of Fig.~\ref{fig:wigner_dynamics}(a) and Fig.~\ref{fig:wigner_dynamics}(b) contain the correlation dynamics of $G^{(1)}_i(\tau)$ and $G^{(2)}_{ii}(\tau)$, respectively, while the panels in Figs.~\ref{fig:wigner_dynamics}(c)--\ref{fig:wigner_dynamics}(e) depict $W_{\Omega_i}(q,p)$ at the initial time $\tau=0$ and at the tautochrone time $\tau_f=\pi/2$.

The first configuration shows that the Wigner negativity of a Fock state appears in the convergence window even when the second-order correlation vanishes. As shown in Fig.~\ref{fig:wigner_dynamics}(a1) we first examine the trajectory of the wavepacket through the density $G^{(1)}_i(\tau)$. A characteristic interference pattern emerges when the Gaussian lobes converge at the tautochrone time. The $G^{(1)}_i$ distribution is identical for the single particle and for classical (coherent-state) light \cite{Kiorpelidis2026}. In the single-particle case the second-order autocorrelator vanishes, $G^{(2)}_{ii}(\tau)=0$ in Fig.~\ref{fig:wigner_dynamics}(b1), since one particle cannot produce a second-order coincidence. The Wigner distributions, by contrast, resolve the particle-number content of each selected window. In particular, in Fig.~\ref{fig:wigner_dynamics}(c1) at $\tau=0$ the state in $\Omega_1$ is the mixture of vacuum and single-particle states $\mathbf{P}(\Omega_1) = (0.5,0.5,0,0)$; the $0.5/0.5$ weighting follows because the delocalized particle carries equal amplitude in $\Omega_1$ and $\Omega_3$, so tracing out the remaining sites leaves a half-vacuum reduced state. At the tautochrone time we have $\mathbf{P}(\Omega_1) = (1,0,0,0)$, which is direct evidence that no population remains in the first observational window. However, in the second window, where $\mathbf{P}(\Omega_2) = (0,1,0,0)$, $W_{\Omega_2}(q,p)$ captures the characteristic negativity and shape of a single-particle Fock state. In the third window, as shown in Fig.~\ref{fig:wigner_dynamics}(e1), the trajectory of the Gaussian lobe is symmetric with respect to the middle site, so the reduced state is identical to that of Fig.~\ref{fig:wigner_dynamics}(c1).

The second configuration produces a clean two-particle Fock state at convergence, despite the two particles originating from separate sites. In Fig.~\ref{fig:wigner_dynamics}(a2) we present this two-particle case, where the mean occupation number shows tautochrone dynamics similar to those of the bosonic case of Fig.~\ref{fig:2particles}(a). The autocorrelation peaks in Fig.~\ref{fig:wigner_dynamics}(b2) are a clear depiction of particles occupying the same spatial region at the tautochrone time. Initially there was no coincidence, because at $\tau=0$ the particles occupy one each of the windows $\Omega_1$ and $\Omega_3$. The form of $W_{\Omega_1}(q,p)$ in Fig.~\ref{fig:wigner_dynamics}(c2) confirms the presence of a single particle, since $\mathbf{P}(\Omega_1) = (0,1,0,0)$. The two particles converge in the tautochrone region at $\tau_f=\pi/2$, where $W_{\Omega_2}(q,p)$ in Fig.~\ref{fig:wigner_dynamics}(d2) shows the Laguerre profile of the two-particle Fock state, with $\mathbf{P}(\Omega_2) = (0,0.01,0.99,0)$. The third window $\Omega_3$ in Fig.~\ref{fig:wigner_dynamics}(e2) is the same as the first window discussed above.

The third configuration shows that a three-particle Fock state still forms at convergence even from an asymmetric, partly co-located initial state. Imposing this configuration, we observe an asymmetry with respect to the middle site in the first-order and second-order correlations in Fig.~\ref{fig:wigner_dynamics}(a3) and Fig.~\ref{fig:wigner_dynamics}(b3), respectively. In the second-order autocorrelation, the two particles that start co-located at site 5 give a nonzero signal along their shared trajectory for all $\tau$, not only at the convergence time. Initially, at $\tau=0$, those two particles form the corresponding $W_{\Omega_1}(q,p)$, as shown in Fig.~\ref{fig:wigner_dynamics}(c3), with $\mathbf{P}(\Omega_1) = (0,0,1,0)$. Here, $P_3(\Omega_1)=0$, because the third particle is located in the third window $\Omega_3$, where $W_{\Omega_3}(q,p)$ has $\mathbf{P}(\Omega_3) = (0,1,0,0)$ as depicted in Fig.~\ref{fig:wigner_dynamics}(e3). At the tautochrone time, however, the three particles converge as shown in $W_{\Omega_2}(q,p)$, which has $\mathbf{P}(\Omega_2) = (0,0,0.01,0.99)$. Thus, despite the asymmetric initial configuration, the selected window realizes an almost pure three-particle Fock state.

Across all cases, the particles measured within the second window at the tautochrone time reproduce, to high fidelity, the profiles of the one-, two-, and three-particle Fock states, as described by the corresponding Laguerre polynomials. This confirms that the particles converge to the tautochrone region. The same Wigner-function analysis can be applied to other input configurations, such as two coherent states, a number state paired with a coherent state, or a squeezed state paired with a coherent state.

\section{Quantum tautochrone with boson interactions}
\label{sec:interactions}

The preceding sections established the tautochrone dynamics for non-interacting particles. We now restrict our study to bosonic atoms and examine how on-site interactions affect the convergence time and the strength of diagonal correlations. In particular, the Hamiltonian becomes:

\begin{equation}
\hat{H}_{\text{total}} = \hat{H} + \frac{U}{2} \sum_{i=1}^L \hat{n}_i (\hat{n}_i - 1)\label{eq:Interactions}
\end{equation}
where $U$ is the strength of the on-site interactions. In cold-atom optical lattices, this on-site interaction sets the energy cost for multiple particles occupying the same lattice site and can be tuned using Feshbach resonances~\cite{Inouye1998}.

\begin{figure}[t]
	\centering
	\includegraphics[width=1\linewidth]{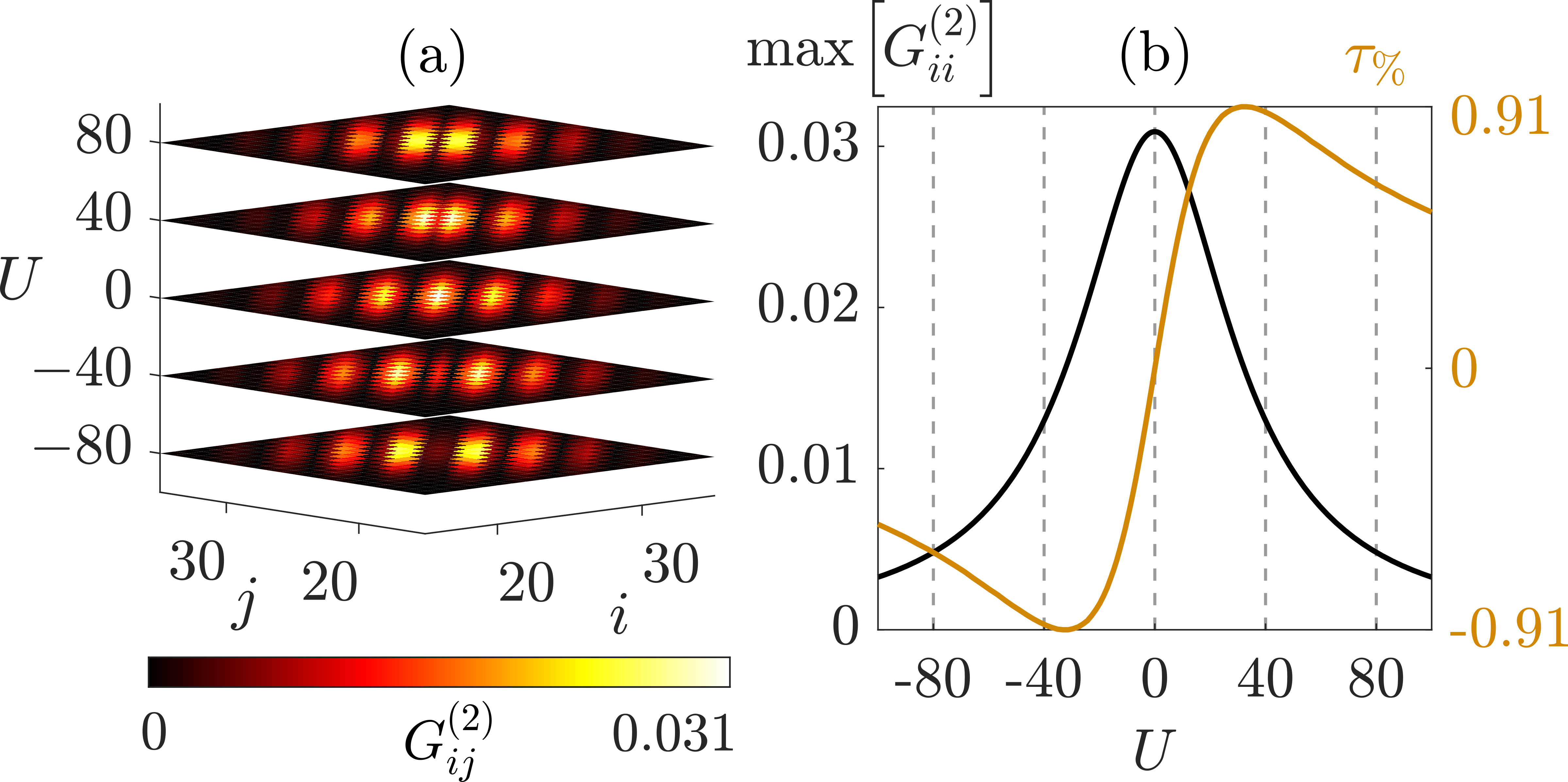}
	\caption{Interaction-dependent two-boson dynamics.
		The initial state consists of two particles centered at sites $8$ and $42$ with $\sigma_m = 2$ considering $L=51$ lattice sites. 
		(a) Snapshots of the full second-order correlator $G^{(2)}_{ij}$, each evaluated at $\tau_f=\pi/2$ for five representative values of $U$. 
		(b) The peak autocorrelation value $\max[G^{(2)}_{ii}(\tau_f=\pi/2)]$ plotted as a function of $U$ (solid black). Relative deviation $\tau_{\%} \equiv \frac{\tau_{\max} - \pi/2}{\pi/2} \times 100\ [\%]$, where $\tau_{\max}$ is the dimensionless time at which the autocorrelator $G^{(2)}_{ii}$ reaches its maximum (solid golden yellow).}
	\label{fig:G1G2_interactions}
\end{figure}

In Figure \ref{fig:G1G2_interactions}, we characterize the dynamics of the two interacting particles at the tautochrone time. As shown in Fig.~\ref{fig:G1G2_interactions}(a), which displays the full second-order correlator $G^{(2)}_{ij}$ evaluated at $\tau_f = \pi/2$, the presence of interactions ($U \neq 0$) alters the spatial distribution of the two-particle correlations. At the tautochrone time, while the non-interacting case exhibits high correlation values near the diagonal in $G^{(2)}_{ij}$, in the presence of interactions those values shift toward off-diagonal elements with $i \neq j$. A clear suppression of the peak autocorrelation, $\max[G^{(2)}_{ii}(\tau_f)]$, as a function of $U$ is evident, as shown in Fig.~\ref{fig:G1G2_interactions}(b).

This suppression can be understood by considering the interplay between interactions and hopping, which is qualitatively different in the weak- and strong-interaction regimes. For weak interactions, it is possible to compensate for the exchange of energy between the hopping term and the interactions term. Thus, the wavepackets converge at the tautochrone. Furthermore, repulsion ($U > 0$) slows the convergence of the wavepackets, delaying the time of the tautochrone, as demonstrated in Fig.~\ref{fig:G1G2_interactions}(b). In contrast, weak attractive interactions ($U < 0$) accelerate the spatial convergence of the particles, leading to an earlier collision time \cite{Valiente2008,Ahlbrecht_2012}. 

For large $|U|$, however, the bound-state band is pushed far above the scattering-state band for $U>0$ and far below it for $U<0$. Since the particles are initially well separated, $\langle \hat H_{\rm int}\rangle \simeq 0$ and $\langle \hat H_{\rm total}\rangle \simeq \langle \hat H\rangle$. As this total energy is conserved, any interaction energy of order $U$ acquired once the wavepackets overlap must be offset by a change in the hopping energy, whose accessible range is limited by the finite bandwidth set by the couplings $J_i$. The initially separated state remains in the scattering-state band, so for the wavepackets to converge, a change in $\langle \hat H\rangle$ beyond this bandwidth would be required \cite{Deuchert2012}. Strong interactions, either attractive or repulsive, suppress second-order autocorrelations and enhance the cross-correlators, as also observed in Fig.~\ref{fig:G1G2_interactions}(a). Consequently, the wavepackets converge only at low values of the interaction strength, while Fig.~\ref{fig:G1G2_interactions}(a) shows that for large values they instead occupy the sites adjacent to the middle site.

\section{Conclusions and perspectives}
\label{sec:conclusions}
The tautochrone effect was recently introduced in the context of classical guided optics \cite{Kiorpelidis2026}. Here we have theoretically examined  the quantum tautochrone effect for both photons in waveguide lattices, and atoms in cold atom lattices. We have showed that the effect persists across a broad class of initial states, provided that the input wavepackets have a finite spatial width. Our analysis demonstrates that both bosonic and fermionic excitations are driven toward the tautochrone region, although their correlation structure remains strongly shaped by the underlying quantum statistics. In particular, bosons can exhibit convergence at the location of the tautochrone, whereas fermions, constrained by the Pauli exclusion principle, approach the same convergence region without occupying the same site. We further showed that the quantum character of the arriving states can be also described by Wigner functions, allowing individual particle-number (Fock) components to be identified at the convergence point. Finally, we have systematically investigated the effect of atomic interactions between bosons to the tautochrone time.

Our results are related  to different classical and quantum physics directions. In particular, from the perspective of quantum optics, the tautochrone dynamics converges photons launched in different waveguides into the same spatial region at a common propagation distance. The resulting multiphoton states may be relevant to multiphoton interference in integrated photonic circuits. A natural question is whether the tautochrone dynamics can lead to the convergence of states of metrological interest, such as squeezed states, and how robust these states are against disorder. From the atomic perspective, since the interaction strength can be tuned experimentally, the effect of weak/strong interactions to the tautochrone effect in a Bose-Hubbard lattice, would be of interest. In this case the finite density of atoms and the associated many-body interactions could play a critical role. 

Beyond engineered lattices, signatures of the quantum tautochrone may also be accessible in many-body dynamics in optical traps without an underlying lattice~\cite{kinoshita_quantum_2006} and in slow-light propagation through elongated atomic ensembles~\cite{Hau1999, Fleischhauer2000, Lukin2003}. 

In conclusion, our results show that nonuniform lattices provide a versatile platform for controllable wavepacket focusing and quantum state engineering.



\section*{Methods}

For the non-interacting calculations, we represented the single-particle hopping Hamiltonian of Eq.~\eqref{eq:Hamiltonian} in the single-particle site basis and propagated each initial wavepacket numerically. The many-particle states were then constructed from these propagated modes with the appropriate bosonic or fermionic exchange symmetry. The correlation functions were evaluated directly using the canonical commutation or anticommutation relations, yielding the expressions presented in the main text. The coupling profile and tautochrone time were obtained from the Heisenberg equations for the lattice position operator, as detailed in Supplementary Note~1. For the interacting calculations, the Bose--Hubbard Hamiltonian was constructed in the complete two-boson Fock basis, including the diagonal on-site interaction term of Eq.~\eqref{eq:Interactions}. All numerical results were obtained by exact unitary evolution, without employing a mean-field approximation. All lattice sizes, initial wavepacket parameters, interaction strengths, and spatial windows are specified in the text or figure captions. For the regional Wigner distributions, the sites outside the selected window $\Omega$ were traced out. The full construction is provided in Supplementary Note~2.

\section*{Data availability}
The data supporting the findings of this study are available from the authors
upon request.

\section*{Code availability}
The numerical codes used to perform the computational analysis reported in
this work are available from the authors upon request.

\begin{acknowledgments}
The authors acknowledge support from the European Research Council (ERC-Consolidator) under Grant Agreement No. 101045135 (Beyond\_Anderson). G. K. S., G. A. S. and K. G. M. further acknowledge funding from the Stavros Niarchos Foundation (SNF) and the Hellenic Foundation for Research and Innovation (HFRI) through the 5th Call of the Science and Society Action, titled Always Strive for Excellence--Theodoros Papazoglou (Project No. 11496, ``PSEUDOTOPPOS''). G. A. S. also acknowledges support from the National Natural Science Foundation of China (NSFC) through Grant No. 12474262.
\end{acknowledgments}

\section*{Author contributions}
G.K.S. and G.E.P. performed the analytical calculations and numerical simulations, while they also analyzed the
results. I.K. supported the reinforcement of theoretical aspects, and analysis of the results. G.A.S. and K.G.M. conceived the idea and supervised the project. All
authors contributed to the discussions and manuscript preparation.	

\section*{Corresponding authors}
Correspondence and requests for materials should be addressed to
Georgios K. Simopoulos, Georgios A. Siviloglou or Konstantinos G. Makris.

\end{document}